\documentclass[traditabstract]{aa} 
\usepackage{graphicx}
\usepackage{txfonts}
\usepackage{natbib,twoopt}
\usepackage{textcomp}
\usepackage{longtable}
\usepackage{booktabs}
\usepackage{lscape}
\usepackage{xspace}
\usepackage{tabularx}
\begin{document}
\title{Circumstellar Disks revealed by $H$\,/\,$K$ Flux Variation Gradients}
\titlerunning{Circumstellar Disks and the FVG method}

\author{
  F. Pozo Nu\~nez
  \inst{1}
  \and
  M. Haas
  \inst{1}
  \and
  R. Chini
  \inst{1,2}
  \and
  M. Ramolla
  \inst{1}
  \and
  C. Westhues
  \inst{1}
  \and
  K.-W. Hodapp
  \inst{3}
}
\institute{
  Astronomisches Institut, Ruhr--Universit\"at Bochum,
  Universit\"atsstra{\ss}e 150, 44801 Bochum, Germany
  \and
  Instituto de Astronom\'{i}a, Universidad Cat\'{o}lica del
  Norte, Avenida Angamos 0610, Casilla
  1280 Antofagasta, Chile
  \and
  Institute for Astronomy, University of Hawaii,
  640 N. Aohoku Place, Hilo HI 96720, USA
}

\authorrunning{F. Pozo Nu\~nez et al.}

\date{Received ; accepted}

\abstract{
The variability of young stellar objects (YSO) changes their
brightness and color preventing a proper classification in traditional
color-color and color magnitude diagrams. We have explored the
feasibility of the flux variation gradient (FVG) method for YSOs,
using $H$ and $K$ band monitoring data of the star forming region
RCW\,38 obtained at the University Observatory Bochum in Chile.
Simultaneous multi-epoch flux measurements follow a linear relation
$F_{H}=\alpha + \beta \cdot F_{K}$ for almost all YSOs with large
variability amplitude. The slope $\beta$ gives the mean $HK$ color
temperature $T_{var}$ of the varying component.
Because $T_{var}$ is hotter than the dust sublimation temperature,
we have tentatively assigned it to stellar variations.
If the gradient does not meet the origin of the flux-flux
diagram, an additional non- or less-varying component may be
required. If the variability amplitude is larger at the shorter wavelength, e.g.
$\alpha < 0$, this component is cooler than the star (e.g. a
circumstellar disk); vice versa,
if $\alpha > 0$, the component is hotter like a
scattering halo or even a companion star. We here present examples of
two YSOs, where the $HK$ FVG implies the presence of a circumstellar
disk; this finding is consistent with additional data at $J$ and $L$.
One YSO shows a clear $K$-band excess in the $JHK$ color-color
diagram, while the significance of a $K$-excess in the other YSO
depends on the measurement epoch. Disentangling the contributions of
star and disk it turns out that the two YSOs have huge variability
amplitudes ($\sim 3-5$\,mag).
The $HK$ FVG analysis is a powerful complementary tool to analyze the
varying components of YSOs and worth further exploration of monitoring
data at other wavelengths.
}

\maketitle
%

\section{Introduction}

The paradigm of a low-mass young stellar object (YSO) involves a
pre-main-sequence star surrounded by a circumstellar disk (CSD), a
(bi)-polar reflection nebula and an envelope. Establishing the
presence of a CSD is an observational challenge.

The classical technique is to reveal the presence of ($T \sim
1500$\,K) dust of the CSD by means of a $K$-excess in $JHK$
color-color diagrams; the YSO is then located at large $H-K$ values
compared to small $J-H$, i.e. right-hand of the reddening vector
(e.g. \citealt{1997AJ....114..288M}). One problem of this technique is
to distinguish between the various emission components viz. star,
disk, and a possible reflection nebula. When the CSD is seen nearly
edge-on, the star is dimmed and heavily reddened so that the apparent
stellar temperature approaches that of the CSD dust. Likewise only a
minor contribution from the less reddened reflection nebula may be
recognisable. In the $JHK$ diagram such a type-2 YSO will be located
at both large $J-H$ and large $H-K$ values, hence only barely
distinguishable from highly reddened sources without CSDs
(e.g. \citealt{1993ApJ...414..676K}, \citealt{2006ApJ...649..900S}).

Instead of analysing multi-component YSOs in magnitude units, it
appears more promising to consider flux (density) or energy
units. Meanwhile numerous multi-wavelength spectral energy
distributions (SEDs) are available, but mostly observed at different
epochs; this limits potentially the results of SED analyses in case of
variable objects. However, particularly YSOs undergo strong brightness
variations at X-ray, optical and near infrared (NIR) wavelengths, for instance
due to stochastic fluctuations of the accretion or to rotating
hot/cool spots on the stellar surface.

Therefore a valuable complement to multi-wavelength SED studies is the
variability monitoring of YSOs, preferably in the NIR to be
less sensitive to extinction. Simultaneous data in two or more filters
provide valuable color information. Assuming that the variations
originate in the star and not in the circumstellar matter, then -- in
a given aperture --  one measures the superposition of a varying hot
stellar component and non-varying cool disk component.

This is exactly the analog of what the two-filter flux variation
gradient (FVG) method reveals in Seyfert galaxies: a hot varying
nucleus and a cool non-varying contamination by the host galaxy
(\citealt{1981AcA....31..293C}, \citealt{1992MNRAS.257..659W},
\citealt{2004MNRAS.350.1049G}, \citealt{2010ApJ...711..461S},
\citealt{2011A&A...535A..73H}, \citealt{2012A&A...545A..84P},
2013, 2014, 2015).
If the central source is sufficiently hot to allow for a
Rayleigh-Jeans approximation, the flux is proportional to the
temperature, and therefore a linear relation between fluxes from the
two filters is expected.

To separate the contributions from variable hot stars and a more
constant circumstellar environment, we have explored the capability of
the FVG method for YSOs, using $H$ and $K$ band monitoring data of the
star forming region RCW\,38 obtained at the University Observatory
Bochum in Chile. Here, we report on a first feasibility study, with
the focus on the linear relationship between the $H$ and $K$ fluxes,
and subsequent potential applications.

\section{Near-infrared data}

\begin{figure}
  \centering
  \includegraphics[width=9cm]{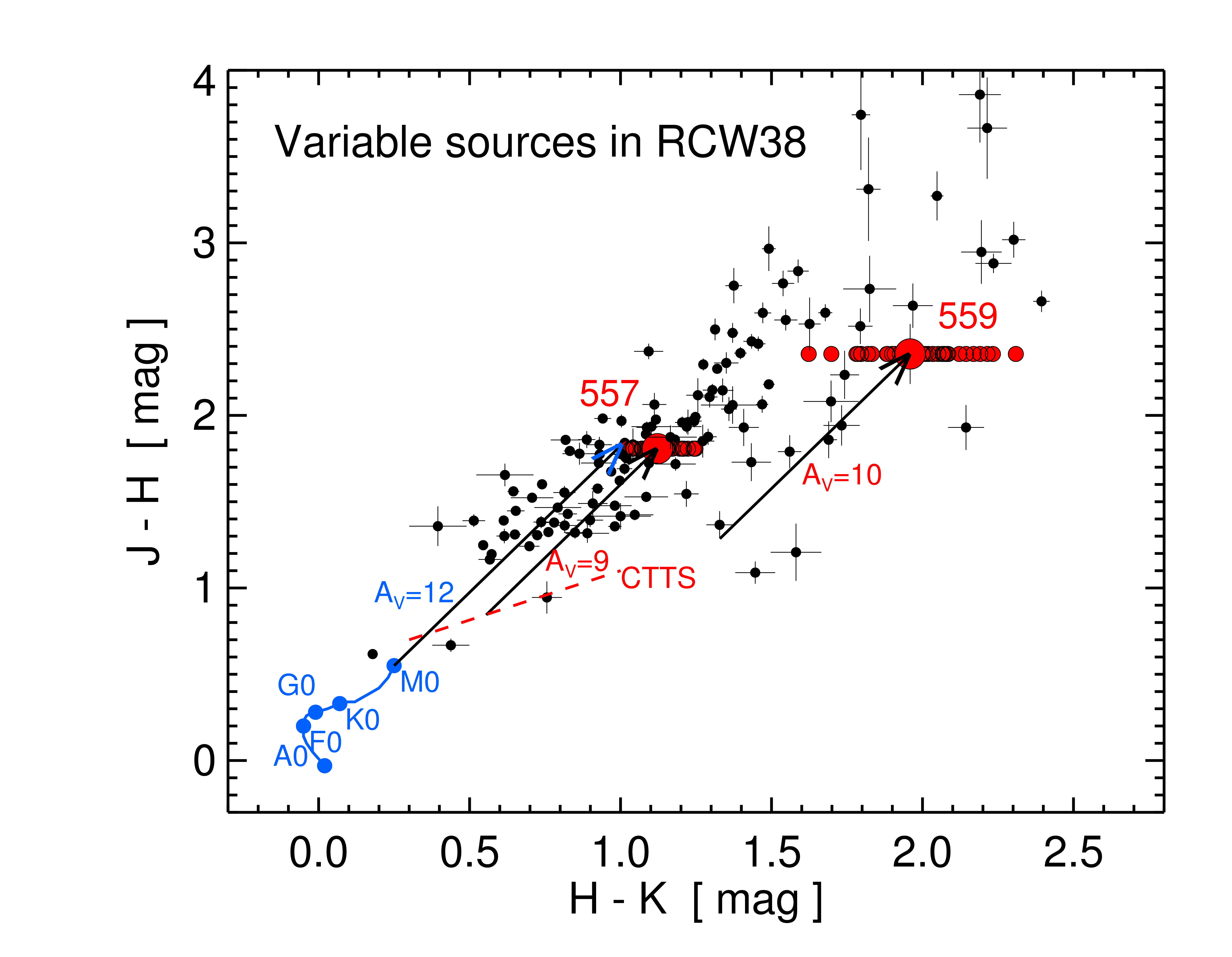}
  \caption{
    Color-color diagram of variable sources in RCW\,38. Black dots
    with error bars denote averaged photometric single observations
    from D\"orr et al (2013). The main sequence is marked in blue and
    the locus of unreddened classical TTauri stars (CTTS) by the red
    long-dashed line. Two sources with ID\,557 and 559 are highlighted
    in red; the small circles mark the individual $H$ and $K$
    measurements placed at mean $J-H$ due to the lack of $J$
    variability data; the large circles give the mean $H-K$.
    The $A_{V}$ vectors are based on \cite{1985ApJ...288..618R}.
  }
  \label{fig_cc}
\end{figure}

The observations, data reduction and PSF photometry have been
described and tabulated by \cite{2013A&A...553A..48D}.
In brief, RCW\,38 has been monitored with the 80\,cm IRIS telescope
(\citealt{2010SPIE.7735E..1AH}) for three months in spring 2011 with a
median daily cadence
in the $J,H,K_s$ bands (henceforth for short $K_s$ is denoted as
$K$). The $H$ and $K$ band light curves
are of good photometric quality; however, the bright nebula causes
a too large uncertainty of the $J$ band
light curves preventing a robust $J$ band variability analysis.

To analyse the mean SED of each YSO, we used the photometry extracted
from the coadded $JHK$ images, reaching about $1-2$\,mag deeper than
2MASS, and complemented the SEDs with Spitzer/IRAC photometry at 3.6
$\mu$m from \cite{2011ApJ...743..166W}.

Fig.~\ref{fig_cc} shows the $JHK$ color-color diagram of 122 variable
sources with $K < 15$\,mag and having
$H$ and $K$ errors smaller than 0.1\,mag (Table\,1 of D\"orr et
al. 2013). For all but two stars the averaged photometric values are
plotted. Two stars (ID\,557 = 2MASS J08591359-4733087 and ID\,559 =
2MASS J08590708-4731499) are selected for a dedicated FVG analysis
here. The small red circles mark their individual $H$ and $K$
measurements. They demonstrate how far a variable source may spread
along the $H-K$ axis. If $J$ variability data were available, one
expects them to increase the spread of the individual measurements
across the $J-H$ vs $H-K$ plane. This demands caution when
classifying YSOs with single epoch data. While ID\,559 shows a strong
$K$ band excess at all epochs, the evidence for a circumstellar disk
around ID\,557 depends on the measurement epoch. If observed only at
an epoch of small $H-K$, ID\,557 could be misidentified as a
main-sequence M0 star reddened by $A_{V} = 12$\,mag, instead of a CTTS
with mean $A_{V} = 9$\,mag.

Our sources ID\,557 and ID\,559 correspond to Winston et al.'s sources
number 363 and 325, respectively. For both sources, the inclusion of
the Spitzer photometry argues in favour of an infrared excess due to
circumstellar dust; Winston et al. assigned "Class\,II" to ID\,557 and
"Flat-Spectrum" to ID\,559. In addition, the $HK$ light curves are
chaotic with jumps between days suggesting that irregular accretion
causes the brightness variations via hot spots, and this in turn
argues for the presence of a circumstellar disk supplying the
accretion material.

\section{$H$\,/\,$K$  Flux Variation Gradients}
\label{section_fvg}

\begin{figure}
  \centering
  \includegraphics[width=8.5cm]{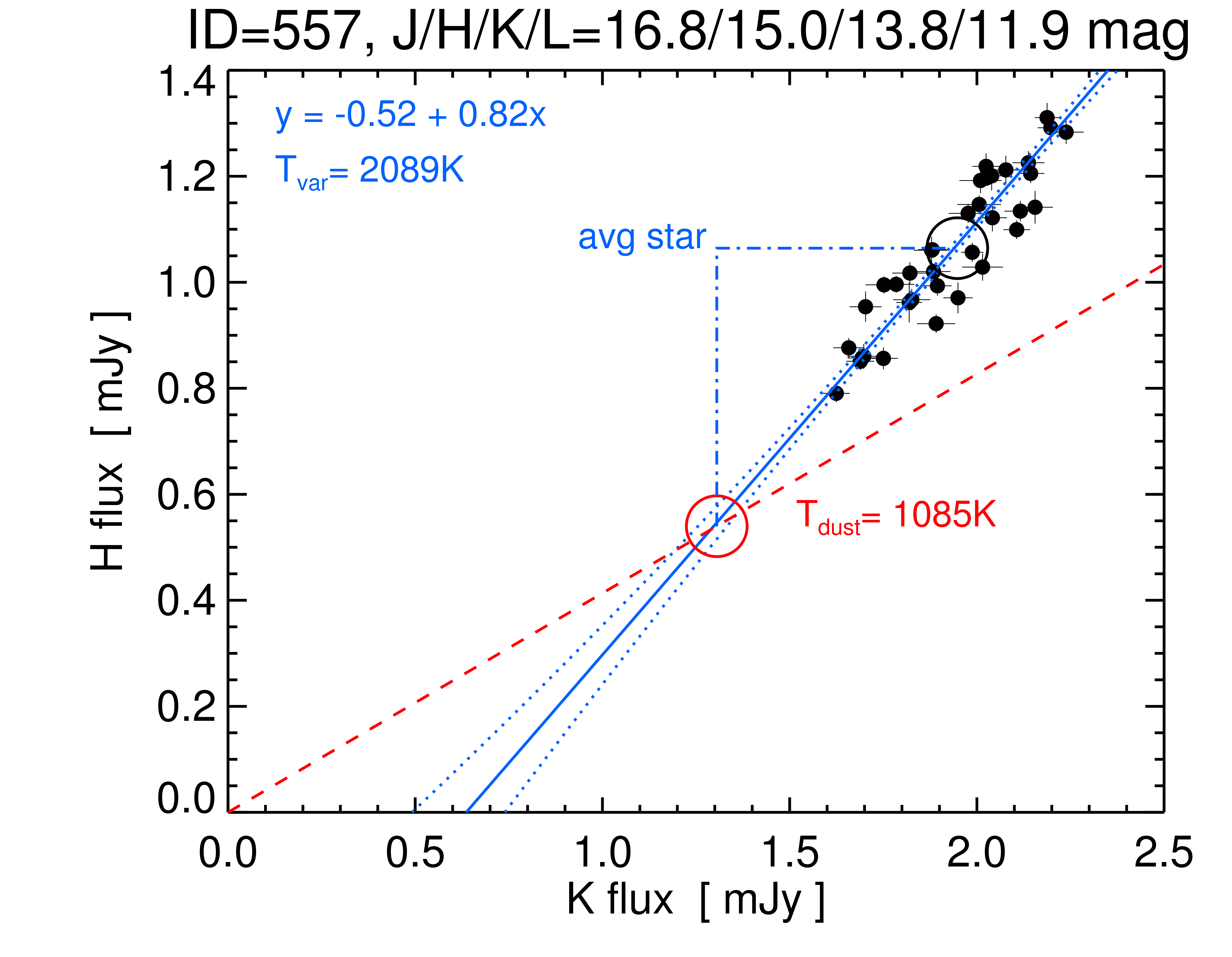}
  \includegraphics[width=8.5cm]{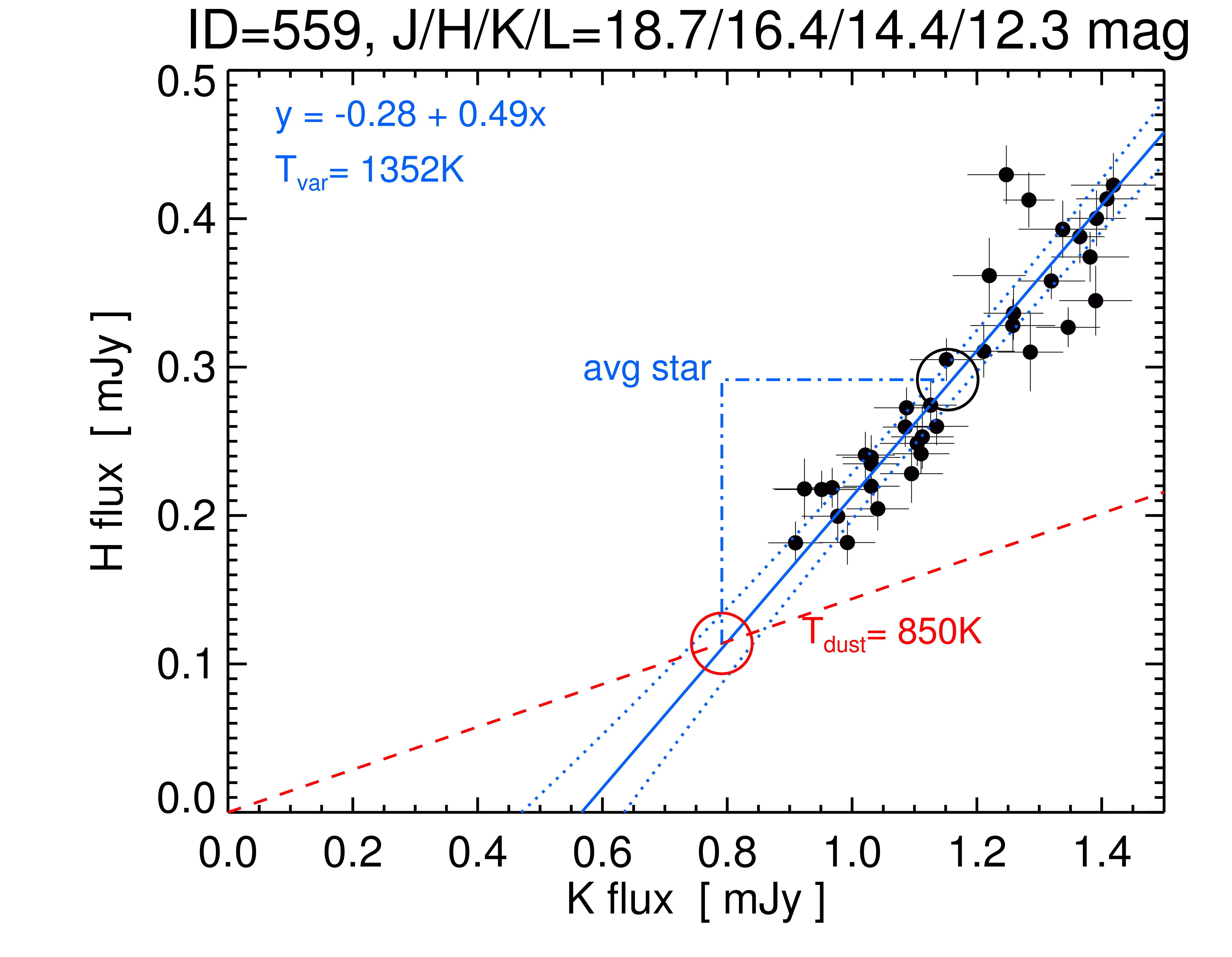}
  \caption{
    Flux - flux diagram of quasi-simultaneous $H$ vs. $K$ measurements
    taken with a time separation of less than two hours (black dots
    with error bars). The data are linearly fit by the blue solid
    line, with an uncertainty range indicated by the dotted blue
    lines. The fit parameters are labeled in the upper left
    corner. The red long-dashed line and the red circle mark the
    constraints for an additional non-varying circumstellar disk,
    derived from the color temperature $T_{var}$ of the varying star
    and the mean SEDs shown in Fig.~\ref{fig_sed}. The blue
    dash-dotted lines mark the average $H$ and $K$ flux of the star,
    given by the difference between the black and red circle.
  }
  \label{fig_fvg}
\end{figure}

Fig.~\ref{fig_fvg} shows the flux - flux diagram for the two sources
ID\,557 and ID\,559. In this diagram, for each (nearly) simultaneous
$HK$ observation $F_{H}$ is plotted versus $F_{K}$ (black dots with
error bars).
The variability data clearly follow a linear relation. The blue solid
line marks the slope $\beta$ of the fitted linear relation
$F_{H}=\alpha + \beta \cdot F_{K}$; this is the flux variation
gradient (FVG). The blue dotted lines delineate the fit uncertainty of
the gradient. The scatter of the data around the gradient is larger
than expected from the error bars. An explanation could be that the
$H$ and $K$ observations were obtained with a lag of about one hour --
hence not sufficiently simultaneous given the strong variations from
one day to the next.

The existence of a linear $HK$ flux relationship shows that the
variable component is intrinsically (i.e. before reddening) hot enough
to be approximated by the Rayleigh-Jeans tail of a black body. The FVG
slope gives to first order the $HK$ color temperature, $T_{var}$, of
the variable component. The apparent $T_{var}$ values of 1352\,K and 2089\,K
(Fig.~\ref{fig_fvg}) are obviously too
low for a stellar source. However, they turn into physically
meaningful stellar temperature
ranges  when applying proper dereddening (see
Sect. \ref{section_deredden}).
Then $T_{var}$ lies well above the expected sublimation temperature
of dust ($T_{dust}$\,$<$\,2000\,K), rejecting the possibility
that the variations are dominated by temperature changes of the dust.

{\it To facilitate the illustration of the FVG method}, we here
tentatively assume that the varying component is the star
itself peppered with hot spots ($T$\,$\sim$\,8000\,K); variations of
the CSD, the reflection nebula or the envelope
are neglected.
In Sect. \ref{section_discussion} we discuss alternatives
to these assumptions, however, they do not affect the basic conclusions from
the FVG technique.

A striking feature of both YSOs in Fig.~\ref{fig_fvg} is that
the FVG
hits the x-axis at $K$\,$\approx$\,0.5\,mJy and does not pass through
the origin of the diagrams.
Note that this is equivalent to the fact that
the shorter wavelength ($H$) varies with larger amplitude than
  the longer wavelength ($K$).\footnote{The
    amplitudes $A=(max-min)/avg$
    are $A(H)$\,=\,0.49, $A(K)$\,=\,0.32 for ID\,557
    and $A(H)$\,=\,0.84, $A(K)$\,=\,0.44 for ID\,559.
  }
Because the flux variations follow a
linear gradient over a large range, it is unlikely that the gradient
will change to hit the origin, if the star fades below its currently
measured $H$ band minimum. Instead, the flux zero point of the star
might lie somewhere on the gradient, between the lowest measured $H$
flux and $H$\,=\,0. This suggests the presence of an additional
component which is cooler than the varying star.

We here identify this additional component with the CSD; the envelope
is expected to be too cold to contribute at $JHK$. Note that for
ID\,559 the $K$ band excess already implies the presence of such a CSD
component; for ID\,557 the $HKL$ data suggest the presence of warm
dust. The location of the CSD in Fig.~\ref{fig_fvg} is marked with a
red circle, and we justify the CSD color estimates in
Section~\ref{section_sed}. Any CSD variations are either negligible or
on an uncorrelated time scale different from that of the star, as
indicated by the small scatter of the data points around the
gradient.

Not only the two selected sources but almost all sources in RCW\,38
with large variability amplitudes $A$
show a linear relationship between $F_{H}$ and $F_{K}$. About 50\% of
the 122 variable sources have
$A>0.4$. About 90\% of them clearly show a linear
gradient, which for $\sim$80\% of the
sources does not go through the origin of the flux-flux diagram.

\section{Decomposition of the mean $JHKL$ SEDs}
\label{section_sed}

To constrain in Fig.~\ref{fig_fvg} the zero point of the star on the
FVG and the contribution of the inferred CSD, we make use of the mean
SED in $JHK$ and $L$ (i.e. Spitzer/IRAC 3.6 $\mu$m). While the $L$
band data were measured at a different epoch, one may expect that the
CSD outshines any stellar contribution at L, hence that any
variability at $L$ is small compared to $JHK$.

\begin{figure}
  \centering
  \includegraphics[width=8cm]{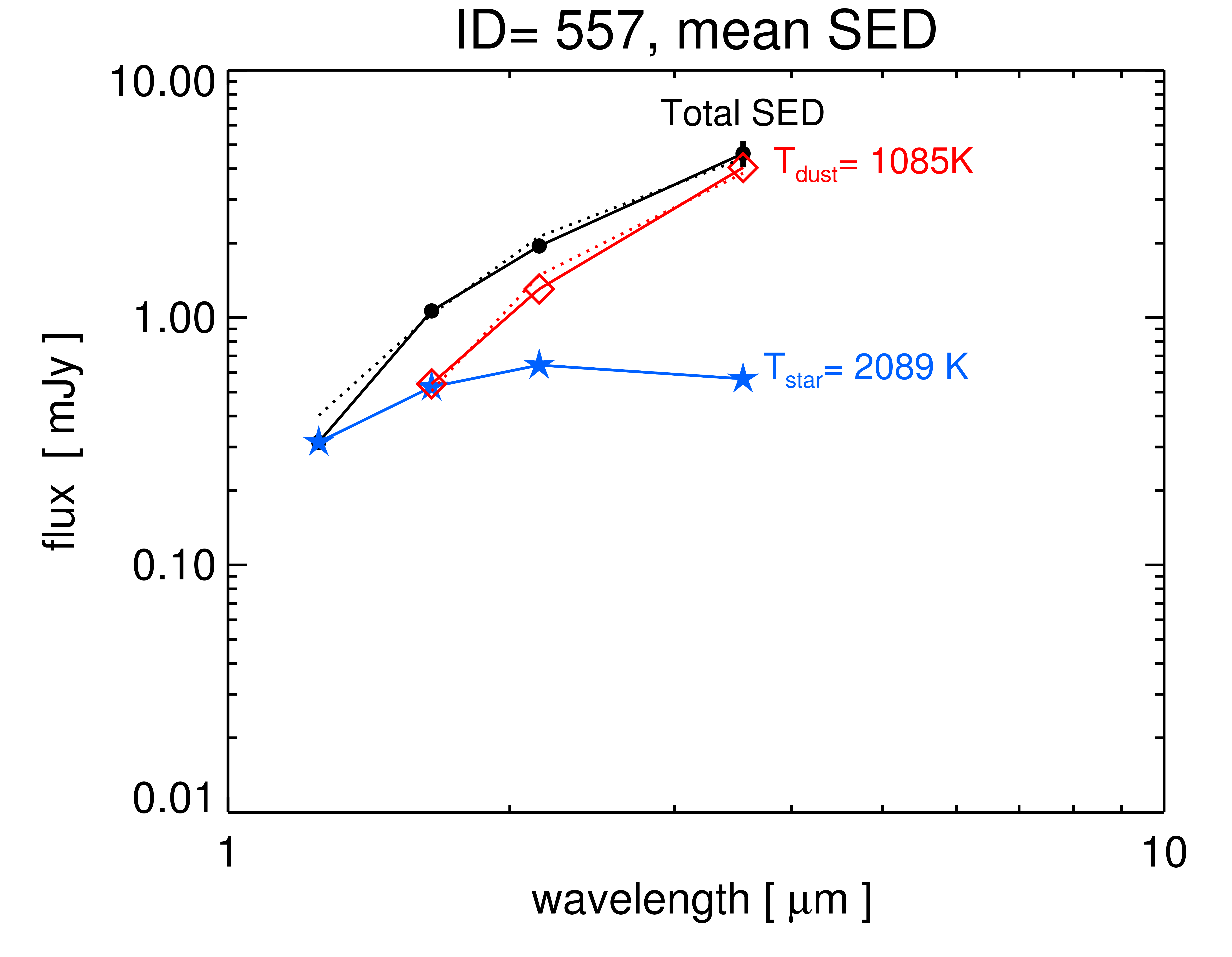}
  \includegraphics[width=8cm]{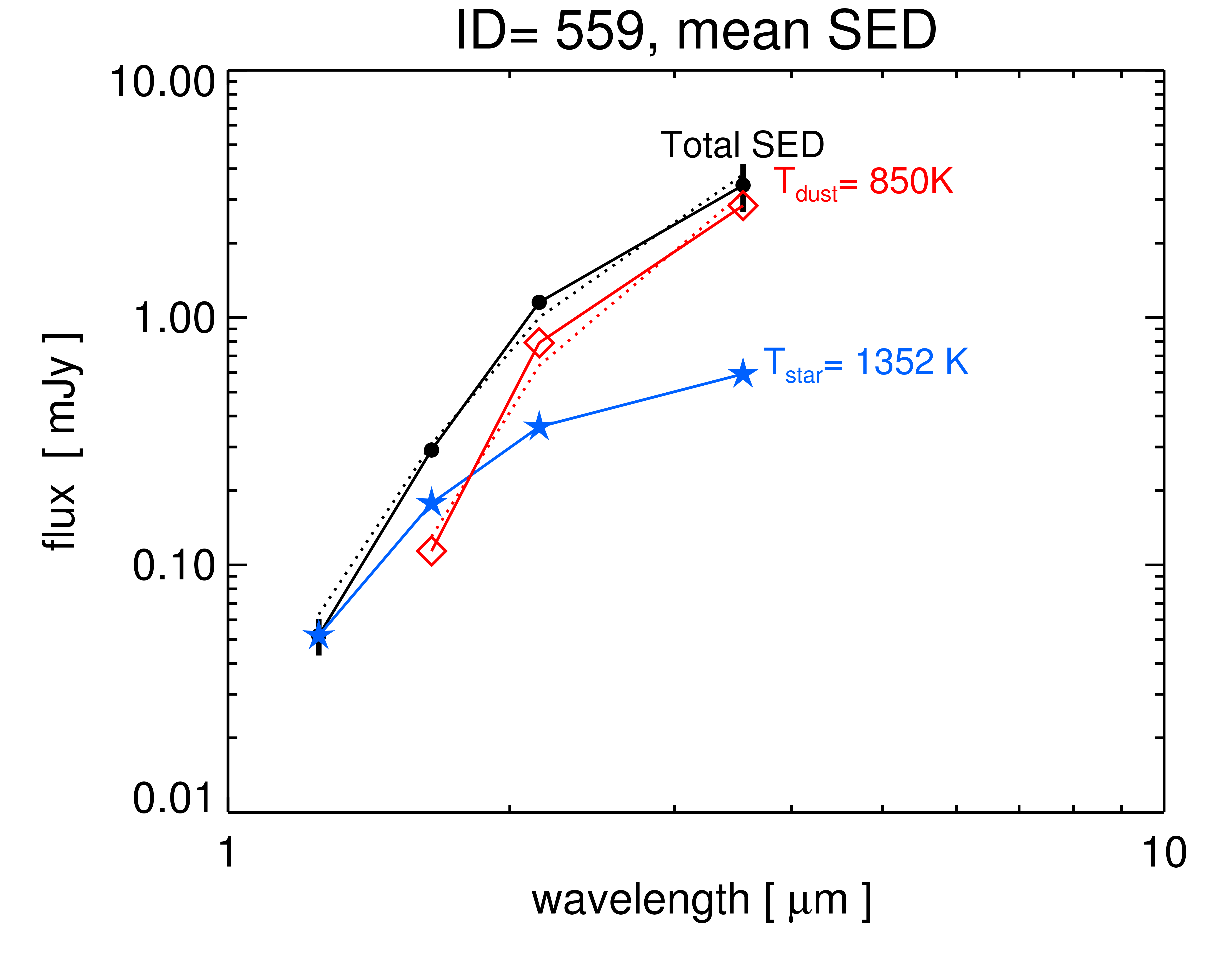}
  \caption{
    Mean Spectral Energy Distributions of ID\,557 and 559. The total
    SED (black) is decomposed into a star (blue) and a circumstellar
    disk (red). The color temperature of the star is identified with
    $T_{var}$ as
    determined from the FVG analysis (Fig.~\ref{fig_fvg}). The star's
    maximum possible mean flux is constrained by the lowest total SED,
    here the $J$ band flux. The remaining SED (total minus star,
    depicted as red diamonds and connected by red solid lines) is
    fitted by a black body (red dotted line). The sum of the star and
    the CSD black body yields the "modeled" total SED (black dotted
    lines).
  }
  \label{fig_sed}
\end{figure}

Fig.~\ref{fig_sed} shows the mean SEDs at $JHKL$. The $HK$ color temperature $T_{var}$ of the star as determined from
the FVG analysis is bluer than that of the total source. For
simplicity we model the star's $JHKL$ SED as a black body with
temperature $T_{var}$. This allows us to constrain the maximum
possible stellar flux, such that it matches the total SED at the $J$
band; in this case we assume that the $J$ band emission of the CSD is negligible.

To check whether the remaining SED (total SED minus "maximum possible"
star) has a reasonable shape consistent with expectations for a CSD,
we fit it by a black body leaving temperature and intensity as free
parameters. As shown in Fig~\ref{fig_sed}, the fit (dotted red line)
deviates from the data (red diamonds and solid lines) by less than
20\%. Furthermore, the sum of the star and the dust black bodies
yields the "modeled" total SED (black dotted lines). Again, the fit at
$HKL$ is good; at $J$ the modeled total SED lies slightly above the
data, but this is expected from setting the star's flux equal to the total $J$
flux.

The resulting mean $HK$ fluxes for the CSDs are plotted as red circles
in Fig.~\ref{fig_fvg} and the maximum possible average star fluxes are
marked with the dashed-dotted blue lines (difference between mean star
flux and mean CSD). The true contribution of the CSD and the star may
be a bit larger/smaller, respectively, because in the above
calculation we have assumed that the CSD does not emit at $J$ and that
the stellar SED follows a black body.

Of course, there may be alternatives to provide constraints on
decomposing the SEDs and to fit a star and a CSD. The investigation of
alternatives is postponed to the future.

\section{Dereddening of star and disk}
\label{section_deredden}

In Fig.~\ref{fig_fvg} the derived $HK$ color temperature of star and
CSD are about $T_{star} \approx 1300, 2100$\,K and $T_{dust} \approx
800, 1100$\,K. This is much lower than the range expected for an
unreddened PMS star, with $T_{star} \approx 3000-8000$\,K and hot dust
($T_{dust} \approx 1500$\,K) in the CSD. Thus, we check if reddening
has "shifted" $T_{star}$ and $T_{dust}$ from reasonable temperatures
to the observed low values.

We assume that both star and CSD are reddened by the same amount of
extinction. Dereddening the JHKL photometry by the $A_{V}$ values from
Fig.~\ref{fig_cc} and the extinction curve from
\cite{1985ApJ...288..618R} we obtain dereddened flux-flux diagrams that look qualitatively
similar to those in Fig.~\ref{fig_fvg} but where the scaling differs
in the sense that $F_{H}$ is more stretched than $F_{K}$. For $A_V =
9$ the dereddened values for ID\,557 turn into $T_{star} \approx
5800$\,K and  $T_{dust} \approx 1350$\,K; an $A_V$ value of 10\,mag
yields $T_{star} \approx 7700$\,K and  $T_{dust} \approx 1400$\,K.

Notably, dereddening of ID\,559 with $A_V$\,=\,10 (Fig.~\ref{fig_cc})
yields $T_{star} \approx 2300$ and $T_{dust} \approx 1100 K$, hence
does not sufficiently raise the star and dust temperatures. However,
increasing the visual extinction to $A_V = 17$\,mag yields $T_{star}
\approx 4900$\,K and $T_{dust} \approx 1350$\,K. In the $JHK$
color-color diagram dereddening with $A_V = 17$ shifts ID\,559 below
the CTTS line (Fig.~\ref{fig_cc}). This "apparent" inconsistency could
be explained by an unresolved reflection nebula diluting the actual
reddening (e.g. \citealt{2006ApJ...649..900S},
\citealt{2009A&A...493..385K}).

\section{Discussion}
\label{section_discussion}

Given the complexity of YSOs, the different components may
contribute in a different manner to the variability observed
across the optical, NIR and MIR wavelength range.
Based on a simultaneous optical ($R$ band) and
MIR (IRAC 3.6-4.5\,$\mu$m) variability study of NGC\,2264,
\cite{2014AJ....147...82C}
present an interesting
discussion on different possible causes of the variability.
For essentially all disk-bearing sources the variations at
both IRAC filters are correlated.
These variations are also correlated with the R band
variations in 40\% of the sources, indicating increased heating of the disk in response to
variable accretion or hot spots on the surface of the star.
For the remaining 60\% of Cody et al.'s sample with
uncorrelated optical and MIR variability,
we propose an alternative
explanation.
Young stars are believed to be born as binaries
or in multiple systems.
If of different mass, the two binary components
evolve differently, resulting in a
blue star and an (infra)-red star.
They may show independent variations,
the blue star dominating
at optical and and the red star at MIR wavelengths.
If such a binary is unresolved,
an object with uncorrelated optical and MIR variability
will be observed.

Swirling dust clouds
(moving across the line of sight towards the star) produce
varying extinction, resulting in a larger variability amplitude
at the shorter
wavelengths (e.g. \citealt{1996AJ....112.2168S}, \citealt{2001AJ....121.3160C}, \citealt{2006ApJ...636..362I}).
Varying extinction may explain the variability behaviour of
UX Orionis stars believed to be seen at grazing angles of the disk
(e.g. \citealt{1994AJ....108.1906H},
\citealt{2000A&A...363..984B}, \citealt{2003ApJ...594L..47D}).
If varying extinction is the dominant mechanism, then in a
$JHK$ color-color diagram or a color-magnitude diagram (CMD)
the varying
data points of each object should be aligned with the
extinction vector.
Such an alignment is only rarely found, e.g. in less than 10\% of
both the Orion
data of Carpenter et al. and our M\,17 data (\citealt{scheyda2010}).
For RCW\,38, light curves are available only at $H$ and $K$,
but not at $J$.
Our two
YSOs (ID\,557 and ID\,559) show a clear elongation
in the two CMDs $H$/($H$-$K$) and $K$/($H$-$K$), which however is
not well aligned with the direction
of the extinction vector.\footnote{
  The resulting ratios $A_H/A_K$ are 1.84\,$\pm$\,0.09 and 2.19\,$\pm$\,0.10
  for ID\,559 and ID\,557, respectively, which is 3.1 $\sigma$
  and 6.2 $\sigma$ apart from the standard extinction law
  ($A_H/A_K$\,=\,1.56, with a
  ratio of total-to-selective extinction $R$\,$\sim$\,3.1,
  \citealt{1985ApJ...288..618R}). A much larger value
  $R$\,$\sim$\,4 (e.g. \citealt{2008ApJ...686..310H})
  would be reqired (locally at the sources)
  to align the $H$ and $K$
  variability of ID\,559 and ID\,557 with the direction
  of the extinction vector.
  Such a large $R$ value
  does not fit the overall consistency
  of the RCW\,38 sources in the $JHK$ color-color diagram
  with the standard extinction law
  (e.g. Fig.\,3 in \citealt{2013A&A...553A..48D}).
}
We suggest that for both sources variable extinction may be present but
plays a minor role.

As shown in Fig.~\ref{fig_fvg}, the CSD contributes by about 50\% to
the total $H$ and $K$ flux; this holds
also for the dereddened data. Thus, after subtraction of the CSD, the
two young stars have a huge variability amplitude
($A \sim 20$ for ID\,557
  and ID\,559 reaches
$A \sim 100$,
equivalent to $\sim$5\,mag).
This  indicates that their luminosity is essentially
powered by accretion bursts. Future studies may provide clues to
whether such strong flux variations can be explained only by
temperature and/or area
changes of hot spots, or whether there are contributions
by a variable hot gaseous disk, in addition to the cooler
circumstellar disk or whether the huge amplitudes are caused by other effects.

  So far, with the aim to facilitate
  the illustration of the FVG method,
  we have neglected variations of
  the CSD, the reflection nebula or the
  envelope.\footnote
  {The assumption that the variations occur also in
    the CSD, implies the existence of a CSD.
    Hence, in this case the FVG
    technique reveals the presence of a CSD.
  }
  Now we address such variations.
  If there is variable heating of the dust disk in response to
  variable accretion or hot spots on the surface of the star,
  and if the dust reacts fast enough
  (e.g. simultaneous within the duration of the $H$ and $K$
  observations),
  then the FVG measures a combination of the hot stellar
  and  the cool dust flux variations.
  Then (after dereddening) $T_{var}$ is not the temperature of the varying
  stellar component ($T_{var}^{stellar}$\,$\sim$\,8000\,K), rather it is
  lowered by the contribution of the varying dust component
  ($T_{var}^{dust}$\,$\sim$\,1500\,K); hence $T_{var}$ underestimates $T_{var}^{stellar}$.
  Accounting for the constant star and dust contributions and
  using $T_{var}$ instead of $T_{var}^{stellar}$
  in the decomposition of the mean SED
  (Fig.\,\ref{fig_sed}) may result in an underestimation
  of the stellar flux at $H$ and $K$, and hence an overestimation
  of the variability amplitude.
  A comprehensive treatment of the many combinations and aspects
  of NIR FVGs for YSOs, including the effects of reflection nebula
  and envelope,
  is planed for the future.

\section{Summary and outlook}
\label{section_outlook}

We have explored to which extent the flux variation gradient (FVG)
technique can be transferred from active galactic nuclei to varying
young stellar objects (YSOs). The results from $HK$ monitoring data of
the star forming region RCW\,38 are:

\begin{itemize}

\item [$\bullet$]
For almost all YSOs with large variability amplitudes the flux
variations follow a linear relation $F_{H}=\alpha + \beta \cdot
F_{K}$. Such a linear relation between fluxes from the two filters is
consistent with the Rayleigh-Jeans approximation. In this case the FVG
slope $\beta$ gives the color temperature
of the varying component.

\item [$\bullet$]
Temperature considerations indicate that the star dominates the flux
variations and that variations of the circumstellar disk/envelope or
the reflection nebula play a minor role.

\item [$\bullet$]
Two examples where the variability amplitudes are larger at
  shorter wavelenghts and where variable extinction might play a minor
  role, are investigated in more detail. The
negative FVG offsets ($\alpha<0$) imply the
presence of an additional cool component
such as a circumstellar disk. At least for some cases the FVG
technique appears to be more sensitive for detecting circumstellar
disks than the $K$-excess in the $JHK$ color color diagram. It appears
worth using FVGs to reexamine the frequency of disks and the evolution
of the disk dispersal in star forming clusters.

\item [$\bullet$]
The low apparent $HK$ color temperatures for the star and the disk, as
inferred from FVGs, can be explained by substantial reddening. The
extinction, required to deredden star and disk to realistic intrinsic
temperatures, is at least as high as inferred from $JHK$ color color
diagrams, consistent with extinction and scattering models.

\item [$\bullet$]
FVGs provide basic input for the decomposition of SEDs. A more
stringent SED decomposition requires further constraints which may be
obtained from additional monitoring in a third filter and subsequent
FVG analysis across several filter pairs.

\item [$\bullet$]
After subtraction of the circumstellar disk, the two young stars
investigated here display a huge
variability amplitude ($\sim 3-5$\,mag).
If the high amplitudes are not pretended by other
effects to be explored, the luminosity of these YSOs
is essentially powered by accretion bursts.

\end{itemize}
To conclude, FVGs are a powerful tool not only to infer the presence
of a circumstellar disk but also to disentangle stellar and disk
parameters with the help of reasonable assumptions.

To give an outlook on further applications, we have tentatively
applied the FVG technique to two data sets:

\begin{itemize}

\item [$\bullet$]
Simultaneous $JHK$ monitoring of the star forming region M\,17 with
the Infrared Service Facility
(IRSF) at the 1.4\,m telescope in Sutherland, South Africa
(\citealt{scheyda2010}) yield for both filter combinations, $JH$ and
$HK$, that the flux variations follow extremely well a linear
relation; $T_{var}$ agrees within the errors. Note that $JK$ is
redundant, being the product of $JH$ and $HK$. The low scatter around
the FVG is probably also a consequence of the perfect simultaneity of
the observations. There are YSOs where the $HK$ FVG with negative
$\alpha$ implies a circumstellar disk; simultaneously the $JH$ FVG has
a positive $\alpha>0$ implying a reflection nebula or -- more
consistent with the unresolved appearance on optical $BVRI$ images -- an
unresolved blue companion star. Since YSOs are suspected to be born in
multiple systems, such a companion is not unexpected, but needs to be
verified. For instance, the young PMS binary
XZ\,Tau consists of a rather evolved blue star and a less evolved
infrared star (\citealt{1990A&A...230L...1H}).

\item [$\bullet$]
The YSOVAR project lists Spitzer/IRAC 3.6 and 4.5\,$\mu$m monitoring
photometry of IC\,1396\,A and Orion
(\citealt{2009ApJ...702.1507M},
\citealt{2011ApJ...733...50M}).
387 Orion sources have amplitudes $A > 0.2$, and
we find that their 3.6 and 4.5\,$\mu$m flux variations follow a linear
relation. For 154 of these sources (40\%) we find $T_{var} > 1500$\,K,
and for 64 sources (16\%) $T_{var} > 2000$\,K, reaching 3500\,K.

Given that the sources are reddened, the variable component of these
sources appears too hot for being dust.
On the other hand, at 3.6 and 4.5\,$\mu$m most of the emission is due to the inner disk and envelope, if present
(e.g. \citealt {2014AJ....148...92R}).To bring both findings into a consistent picture, both stars and disk/envelope may vary simultaneously with
similar strength so that the FVG measures an average $T_{var}$, higher than for dust but slightly lower than for the star.

\end{itemize}
These two data sets ($JHK$ of M\,17, and YSOVAR)
strengthen the suggestion that the FVG technique
opens the door to valuable insights into YSOs which are hard to obtain
otherwise.

\begin{acknowledgements}

  This work is supported by the Nordrhein-Westf\"alische Akademie
  der Wissenschaften und der K\"unste
  in the framework of the academy program of the Federal Republic of
  Germany and the state Nordrhein-Westfalen, by   Deutsche
  Forschungsgemeinschaft (DFG HA3555/12-1) and by Deutsches Zentrum
  f\"ur Luft-und Raumfahrt (DLR 50\,OR\,1106).
  We thank the anonymous referee for constructive comments and careful review of the manuscript.

\end{acknowledgements}

\bibliographystyle{aa} 
\bibliography{fvg_fp}

\begin{thebibliography}{33}
\expandafter\ifx\csname natexlab\endcsname\relax\def\natexlab#1{#1}\fi

\bibitem[Bertout(2000)]{2000A&A...363..984B} Bertout, C.\ 2000, \aap, 363, 984

\bibitem[Carpenter et al.(2001)]{2001AJ....121.3160C} Carpenter, J.~M., 
Hillenbrand, L.~A., \& Skrutskie, M.~F.\ 2001, \aj, 121, 3160 

\bibitem[Choloniewski(1981)]{1981AcA....31..293C} Choloniewski, J.\ 1981, 
\actaa, 31, 293

\bibitem[Cody et al.(2014)]{2014AJ....147...82C} Cody, A.~M., Stauffer, J., 
Baglin, A., et al.\ 2014, \aj, 147, 82 

\bibitem[D{\"o}rr et al.(2013)]{2013A&A...553A..48D} D{\"o}rr, M., Chini, R., Haas, M., Lemke, R., Nurnberger, D.\ 2013, \aap, 553, A48 

\bibitem[Dullemond et al.(2003)]{2003ApJ...594L..47D} Dullemond, C.~P., van 
den Ancker, M.~E., Acke, B., \& van Boekel, R.\ 2003, \apjl, 594, L47 

\bibitem[Glass(2004)]{2004MNRAS.350.1049G} Glass, I.~S.\ 2004, \mnras, 350, 
1049

\bibitem[Haas et al.(1990)]{1990A&A...230L...1H} Haas, M., Leinert, C., \& Zinnecker, H.\ 1990, \aap, 230, L1 

\bibitem[Haas et al.(2011)]{2011A&A...535A..73H} Haas, M., Chini, R., Ramolla, M., et al.\ 2011, \aap, 535, A73 

\bibitem[Herbst et al.(1994)]{1994AJ....108.1906H} Herbst, W., Herbst, 
D.~K., Grossman, E.~J., \& Weinstein, D.\ 1994, \aj, 108, 1906

\bibitem[Hodapp et al.(2010)]{2010SPIE.7735E..1AH} Hodapp, K.~W., Chini, 
R., Reipurth, B., et al.\ 2010, \procspie, 7735, 77351A 

\bibitem[Hoffmeister et al.(2008)]{2008ApJ...686..310H} Hoffmeister, V.~H., 
Chini, R., Scheyda, C.~M., et al.\ 2008, \apj, 686, 310

\bibitem[Indebetouw et al.(2006)]{2006ApJ...636..362I} Indebetouw, R., 
Whitney, B.~A., Johnson, K.~E., \& Wood, K.\ 2006, \apj, 636, 362 

\bibitem[Kenyon et al.(1993)]{1993ApJ...414..676K} Kenyon, S.~J., Calvet, 
N., \& Hartmann, L.\ 1993, \apj, 414, 676

\bibitem[Kr{\"u}gel(2009)]{2009A&A...493..385K} Kr{\"u}gel, E.\ 2009, \aap, 493, 385 

\bibitem[Meyer et al.(1997)]{1997AJ....114..288M} Meyer, M.~R., Calvet, N., 
\& Hillenbrand, L.~A.\ 1997, \aj, 114, 288 

\bibitem[Morales-Calder{\'o}n et al.(2009)]{2009ApJ...702.1507M} 
Morales-Calder{\'o}n, M., Stauffer, J.~R., Rebull, L., et al.\ 2009, \apj, 
702, 1507 

\bibitem[Morales-Calder{\'o}n et al.(2011)]{2011ApJ...733...50M} 
Morales-Calder{\'o}n, M., Stauffer, J.~R., Hillenbrand, L.~A., et al.\ 
2011, \apj, 733, 50 
 
\bibitem[Pozo Nu{\~n}ez et 
al.(2012)]{2012A&A...545A..84P} Pozo Nu{\~n}ez, F., Ramolla, M., Westhues, C., et al.\ 2012, \aap, 545, A84 

\bibitem[Pozo Nu{\~n}ez et 
al.(2013)]{2013A&A...552A...1P} Pozo Nu{\~n}ez, F., Westhues, C., Ramolla, M., et al.\ 2013, \aap, 552, A1 

\bibitem[Pozo Nu{\~n}ez et 
al.(2014)]{2014A&A...561L...8P} Pozo Nu{\~n}ez, F., Haas, M., Chini, R., et al.\ 2014, \aap, 561, L8

\bibitem[Pozo Nu{\~n}ez et 
al.(2015)]{2015A&A...576A..73P} Pozo Nu{\~n}ez, F., Ramolla, M., Westhues, C., et al.\ 2015, \aap, 576, A73 

\bibitem[Rebull et al.(2014)]{2014AJ....148...92R} Rebull, L.~M., Cody, 
A.~M., Covey, K.~R., et al.\ 2014, \aj, 148, 92

\bibitem[Rieke \& Lebofsky(1985)]{1985ApJ...288..618R} Rieke, G.~H., \& Lebofsky, M.~J.\ 1985, \apj, 288, 618 

\bibitem[Sakata et al.(2010)]{2010ApJ...711..461S} Sakata, Y., Minezaki, 
T., Yoshii, Y., et al.\ 2010, \apj, 711, 461

\bibitem[Scheyda (2010)]{scheyda2010}
Scheyda, C. M., 2010, PhD Thesis, Ruhr-Universit\"at Bochum

\bibitem[Skrutskie et al.(1996)]{1996AJ....112.2168S} Skrutskie, M.~F., 
Meyer, M.~R., Whalen, D., \& Hamilton, C.\ 1996, \aj, 112, 2168

\bibitem[Stark et al.(2006)]{2006ApJ...649..900S} Stark, D.~P., Whitney, 
B.~A., Stassun, K., \& Wood, K.\ 2006, \apj, 649, 900

\bibitem[Winkler et al.(1992)]{1992MNRAS.257..659W} Winkler, H., Glass, 
I.~S., van Wyk, F., et al.\ 1992, \mnras, 257, 659 

\bibitem[Winston et al.(2011)]{2011ApJ...743..166W} Winston, E., Wolk, 
S.~J., Bourke, T.~L., et al.\ 2011, \apj, 743, 166 


\end{thebibliography}

\end{document}